\documentstyle[12pt]{article}

\begin{document}

\begin{flushright} {OITS 628}\\
June 1997
\end{flushright}
\vspace*{1cm}

\begin{center} {\Large {\bf Enhanced $J/\psi$ Suppression Due to Gluon
Depletion }}
\vskip .75cm
 {\bf  Rudolph C. Hwa$^1$, J\'{a}n Pi\v{s}\'{u}t$^2$ and
Neva Pi\v{s}\'{u}tov\'a$^2$ }
\vskip.5cm

{$^1$Institute of Theoretical Science and Department of Physics\\
University of Oregon, Eugene, OR 97403-5203, USA\\
\bigskip
$^2$Department of Physics, Comenius University, SK-84215,
Bratislava, Slovakia\\}
\end{center}

\begin{abstract} The nonlinear effect of gluon depletion in the collision of
large nuclei can be large.  It is due to multiple scatterings among comoving
partons initiated by primary scattering of partons in the colliding
nuclei.   The effect can give rise to substantial suppression of
$J/\psi$ production in very large nuclei, even if the linear depletion effect is
insignificant for the collisions of nuclei of smaller sizes.  This mechanism
offers a natural explanation of the enhanced suppression in the Pb-Pb
data recently observed by NA50.
\end{abstract}

In a previous paper \cite{hpp} we have examined the issues involved in
ascribing some aspect of the phenomenon of $J/\psi$ suppression in
heavy-ion collisions
\cite{for} to the depletion of gluons prior to the hard subprocess of $c\bar
c$ production.  What we have found is that the data on the survival
probability
$S$ without the points from Pb-Pb collisions \cite{cl,mg}, by themselves,
cannot distinguish whether the suppression is due to gluon depletion or
hadronic-nuclear absorption.  That is, both mechanisms contribute to an
exponential dependence of $S$ on the effective path length $L$ (or on
log$AB$).  We now consider the enhanced suppression in the Pb-Pb data of
NA50 \cite{mg} and show how gluon depletion can naturally account for it.
Furthermore, it is possible for that to happen even if the ``normal"
suppression in the lighter-ion data is due mainly to the absorption
mechanism with negligible depletion effect.

Many suggestions have been advanced to account for the enhancement of
$J/\psi$ suppression observed in the Pb-Pb collision data
\cite{kl}-\cite{wo}.  They all refer to the absorption processes after the
production of the
$c\bar c$ state.  Our suggestion is concerned with the depletion of gluons
before  the
$gg\rightarrow c\bar c$ subprocess.  The basic idea is  rather intuitive and
can be described qualitatively before we go into the details.  Consider a row
of nucleons in nucleus $A$ colliding with another row in nucleus $B$, and
suppose that the $n_A$th one from the front of the former (call it $a$)
collides with the $n_B$th one in the latter (call it
$b$) in a hard process creating $c\bar c$. The gluon depletion mechanism
discussed in \cite{hpp} takes into account of the loss of gluons in $a$ (due
primarily to
$g\rightarrow q\bar q$) as it goes through $B$ until $gg\rightarrow c\bar
c$ occurs with a gluon in $b$; similarly, the gluons in $b$ are depleted as
$b$ traverses $A$.  We shall refer to this process as linear depletion for
reasons that will become clear below.  What we now want to emphasize is
that a nonlinear depletion process may be even more important.  Such a
process is due to the interaction of the gluons in $a$ with the
slower partons liberated from the 
$n_A-1$ forerunners in $A$ broken by earlier interactions, and likewise $b$
with the partons of the 
$n_B-1$ forerunners in $B$.  In an
imperfect, yet helpful, analogy one may think of a multicar accident on a
busy, foggy highway and recognize that most of the collisions are between
cars originally going in the same direction.

Let us first summarize the essence of the linear effect considered in
\cite{hpp}.  Our notation
 will follow that of \cite{hpp}, but abbreviated.  The probability that a
nucleon in
$A$ makes
$\nu_1$ collisions in $B$ before the hard subprocess is
\begin{eqnarray}
\pi_{\nu_1}={1\over \nu_1!}{n_B}^{\nu_1}
e^{-n_{_B}}
\quad ,
\qquad n_B=\sigma_{\rm in} T_B^-\quad ,
\label{1}
\end{eqnarray}
where  $T_B^-$ is the path length that is traversed in $B$ before the $c\bar
c$ production and is dependent on the impact parameter $b_B$ and
longitudinal position $z_B$, both being suppressed.  If the depletion factor
per collision at fixed momentum fractions $x_1$ and $x_2$ is $D$ ($D=1$ for
no depletion), then the suppression factor at fixed $b$ and $z$ in $A$ and
$B$ is
\begin{eqnarray}
\Gamma_{AB}^{(d)} =
\sum_{\nu_1,\nu_2}\pi_{\nu_1}\pi_{\nu_2}D^{\nu_1+\nu_2} = {\rm exp}\,
[-(1-D)(n_A+n_B)]\quad ,
\label{2}
\end{eqnarray}
where $\pi_{\nu_2}$ is defined
as in (\ref{1}), but with $B$ replaced by $A$.  It is the simple sum,
$n_A+n_B$, in (\ref{2}) that leads us to call the effect linear.  The
exponential behavior of
$\Gamma_{AB}^{(d)}$ is what generates, after integration over $b$ and
$z$, the  approximate exponential dependence of $S$ on $L$ that is
indistinguishable from the Gerschel-H\"ufner formula \cite{gh}, derived
from purely absorptive consideration.

The nonlinear effect that we now describe arises from the interactions with
the forerunners.  At the partonic level the linear effect is due to the
primary interactions of gluons in nucleons going in opposite directions,
while the nonlinear effect is due to the secondary, tertiary, etc.\
interactions among partons moving in the same direction, initiated by
primary interactions.  The rapidity separation,
$\Delta y$, between the participants of the primary interaction is large
because they belong to the nuclei $A$ and $B$ separately.  On the other
hand,
$\Delta y$ between the partons involved in the secondary (or tertiary,
etc.) interactions is small because they belong to the same nucleus.
Ordinarily, in an unperturbed nucleus or in deep inelastic scattering of a
nucleus, those partons in different nucleons do not interact except in the
context of nuclear binding and shadowing.  However, if a primary interaction
has taken place between two colliding nuclei, the scattered parton in $A$,
whether at large or small angle, can interact with a parton coming from 
behind in the same or neighboring rows.  Since they are comovers, their
interaction can be much stronger than the primary interaction, a property
that is consistent with the general notion of strong interaction  in soft
processes being short-ranged  (in rapidity).
  Thus even if the linear depletion effect is small, the nonlinear effect need
not be.

If we consider an $n_A\times n_B$ matrix, representing the possible
pairings of
$n_A$ and $n_B$ nucleons in collisions, the last row and last column
contribute to the linear depletion effect.  [Their sum $n_A+n_B-1$ appears
as
$n_A+n_B$ in (\ref{2}) in compensation for the fact that the first
collision of a nucleon with a row of nucleons is the normal $pp$ collision,
whose cross section is larger than those of the subsequent collisions that
involve the broken nucleon propagating downstream.  To elaborate on this
point is too much of a digression that is not germane to the following
discussion.] The remaining part of the matrix having $(n_A-1)(n_B-1)$
pairings contributes to the quadratic depletion effect due to multiple
parton scatterings.  Let us define
\begin{eqnarray}
n'_A = (n_A-1)\ \Theta(n_A-1) \quad,
\label{3}\end{eqnarray}
and similarly for
$n'_B$.  Then, assuming $A\leq B$, the average number of collisions that
the forerunners of $a$ in $A$ make with the forerunners of $b$ in $B$,
producing comoving partons that can interact with the partons of $a$, is
$n'_An'_B-{n'_A}^2/2$;  that for producing comoving partons with the
ones in $b$ is
${n'_A}^2/2$.  This way of partitioning the $n'_An'_B$ pairings can be
visualized in the forward light-cone of $AB$ collision, where the former lie
on the $A$ side of the interaction region, while the latter lie on the $B$
side.   The precise method of partitioning is unimportant, as will become
evident presently.

The probabilities that $a$ and $b$ can interact $\nu'_1$ and $\nu'_2$ times
with their respective forerunners are
\begin{eqnarray}
\pi'_{\nu'_1}={1\over \nu'_1!}\ (n'_An'_B-{n'_A}^2/2)^{\nu'_1}\
e^{-(n'_An'_B-{n'_A}^2/2)} \quad,
\label{4}\end{eqnarray}
\begin{eqnarray}
\pi'_{\nu'_2}={1\over \nu'_2!}\ ({n'_A}^2/2)^{\nu'_2}\  e^{-{n'_A}^2/2}
\quad .
 \label{5}\end{eqnarray}
If $D'$ is the effective gluon depletion factor for
each of those interactions, then the corresponding suppression factor,
analogous to (\ref{2}), is
\begin{eqnarray}
\Gamma'^{(d)}_{AB} =
\sum_{\nu'_1,\nu'_2}\pi'_{\nu'_1}\pi'_{\nu'_2}\  D'^{^{\nu'_1+\nu'_2}} =
{\rm exp}\ [-(1-D')\ n'_A\,n'_B]\quad .
\label{6}\end{eqnarray}
We refer to this as the quadratic depletion effect, since it is $n'_An'_B$ that
appears in the exponent, as opposed to $n_A+n_B$ in (\ref{2}).  As it is in
(\ref{2}), the dependences on
$b_A, z_A, b_B$, and $ z_B$ have been suppressed in (\ref{6}).

The combined suppression factor due to both linear and quadratic
depletion as well as absorption \cite{hpp} is now
\begin{eqnarray}
P={\rm exp}\ [-(1-D)\,
(n_A+n_B)-(1-D')\,n'_An'_B-\sigma_a(T^+_A+T^+_B)] \quad,
\label{7}\end{eqnarray}
where $\sigma_a$ is the absorption
cross section and
$T^+_A$ is the path length in
$A$ traversed by the $J/\psi$ system.  Exhibiting the $b$ and $z$
dependences, we have \cite{hpp}
\begin{eqnarray}
T^{\pm}_A=(1-{1\over A})\ \rho_0\ (L_A\pm z_A)\quad,
\qquad\quad L_A=(R^2_A - s^2)^{1/2}\quad,
\label{8}\end{eqnarray}
 and
similarly for
$T^{\pm}_B$, with $\vec b_A=\vec s$ and $\vec b_B=\vec b-\vec s$.  The
average overall suppression factor (more precisely, survival probability)
is
\begin{eqnarray}
S^{AB}_{J/\psi} = N^{-1}_{AB} \int d^2 b \int d^2 s
\int_{-L_A}^{L_A} dz_A
\int_{-L_B}^{L_B} dz_B\ P \quad,
\label{9}\end{eqnarray}
where $N_{AB}$ is the
same integral as in (\ref{9}) but with $P$ replaced by 1.

To see how $S^{AB}_{J/\psi}$ depends on $A$ and $B$, let us examine the
parameters in the formula.  Without the quadratic depletion terms in
(\ref{7}), we have
\begin{eqnarray}
P_1\equiv P(D'=1) = {\rm exp}\ [-\sigma _d\ (T^-_A+T^-_B) - \sigma _a\
(T^+_A+T^+_B)]
\quad,
\label{10}\end{eqnarray}
where $\sigma_d=\sigma_{\rm in}(1-D)$.  As
pointed out in \cite{hpp}, (\ref{10}) exhibits the symmetry between the
depletion effect before the formation of $J/\psi$ and the absorption effect
afterwards.  That is why the exponential dependence of the empirical
$S^{AB}_{J/\psi}$ on the effective length $L$ (or log$AB$) cannot
distinguish the two effects.  So long as the combined cross section
$\sigma_c=\sigma_a+\sigma_d$ is around 7 mb, the heavy-ion data,
excluding the Pb-Pb collisions, can be fitted by any ratio
$\eta=\sigma_d/\sigma_a$.  Now, we consider the contribution from the
quadratic depletion term in (\ref{7}) only, giving
\begin{eqnarray}
P_2 \equiv P(D=1,\sigma_a=0)= {\rm exp} \left[-\ \tau\
(n_A-1)\,(n_B-1)\,\Theta (n_A-1)
\,\Theta (n_B-1) \right]
\label{11}\end{eqnarray}
where $\tau=1-D'$, a parametrization, like $\sigma_d$, having the more
proper sense of depletion in that $\tau=0$ means no depletion.  There are
a number of features of (\ref{11}) worth noting.

\noindent (a) \quad While the discussions in the introduction and in the
paragraph containing (\ref{3}) regard $n_A$ and $n_B$ as integers for the
sake of ease in describing the nonlinear depletion mechanism, they can in
reality have any positive value by virtue of their definitions,
$n_{A,B}=\sigma_{\rm in} T_{A,B}^-$.  That is why the step functions in
(\ref{3}) and (\ref{11}) are important to ensure that the participants of
the process,
$n'_A$ and $n'_B$, are nonnegative.  As a consequence there is a threshold
effect, i.e.,
$A$ and $B$ must be large enough for the  mechanism to be operative.

\noindent (b) \quad The inelastic cross section $\sigma_{\rm in}$ is
relevant in the determination of the position of the threshold.  It is not the
$\sigma^{pp}_{\rm in}$ for $pp$ collision because, except for the first
collisions on the front sides of the nuclei, most of the collisions are between
broken nucleons \cite{hw}, which consist mainly of the parton fluxes that
propagate downstream after the bound nucleons are broken by the first
collisions.
$\sigma_{\rm in}$ is an effective cross section for the collision of such
broken nucleons, and there exist no reliable estimates for its value.  Using
$p'$ to denote broken nucleon, and taking $\sigma^{pp}_{\rm in} \approx
30$ mb, it is not unreasonable to consider $\sigma^{p'p}_{\rm in} \approx
20$-25 mb, and
$\sigma^{p'p'}_{\rm in}\approx (\sigma^{p'p}_{\rm in})^2/
\sigma^{pp}_{\rm in}
\approx 13$-21 mb.  We shall adopt $\sigma_{\rm in}\approx 15$-25 mb as
typical values.

\noindent (c) \quad The quadratic depletion parameter $\tau$ can be
substantially different from zero, even if the linear effect measured by
$\sigma_d$ is zero, since, as discussed earlier, the interaction between
partons with small rapidity separation can be much greater than that
between partons with large $\Delta y$.  Since the determination of $\tau$
from first principles is difficult, we shall use it as a free parameter in the
following.  It should be noted that even if
$D'=0$, i.e., total depletion per collision, $\tau$ attains its maximum value 1,
so (\ref{11}) does not give $P_2=0$. That is because the Poissonian
fluctuations in (\ref{4}) and (\ref{5}) allow for $\nu'_1=\nu'_2=0$,  which
result in a nonvanishing probability for the passage of the gluon fluxes
with minimal influence by the depletion mechanism.

To gain some further insight in the quadratic depletion effect, let us
compute
$S^{AB}_{J/\psi}$, taking only $P_2$ into account, i.e., by substituting
(\ref{11}) alone into (\ref{9}).  Using the integration procedure developed
in
\cite{hpp}, we obtain the results shown in Fig.\ 1, where $\sigma_{\rm in}$
is set at 15, 20 and 25 mb; the shaded regions are bounded by $\tau=0.5$
from above (for illustrative purpose) and $\tau=1.0$ from below.     Evidently, the
threshold for the quadratic depletion effect is higher at smaller
$\sigma_{\rm in}$, since there would be less participants for the
multiscattering subprocesses unless $A$ is higher.  Furthermore, even at
maximum depletion $(\tau=1)$ there is still a residual rate of
$J/\psi$ production because of the aforementioned probability of  gluon
passage without depletion.  We note that, although the parameters
$\sigma_{\rm in}$ and $\tau$ are not empirically familiar, their values used
in Fig.\ 1 are sensible estimates, so the suppression effect revealed is a
natural consequence of a physical process that is not contrived to explain
the data.

For a comparison with the data  \cite{cl,mg} we include both $P_1$ and
$P_2$ in (\ref{9}) and calculate the overall suppression factor.  Since, as
found in \cite{hpp}, the combined effect of absorption and linear depletion
is insensitive to the ratio
$\eta=\sigma_d/\sigma_a$, we choose the uncontroversial values:
$\sigma_c=\sigma_a+\sigma_d=7$ mb, and
$\eta\approx 0.1$.  
For quadratic depletion  effect we use $\sigma_{\rm in}=20$ mb
and
$\tau=0.5$-1.0.  The result is shown by the triangles in Fig.\ 2.  The
agreement with the data [4] is evidently very good.  Of course, if there
exists enhanced nuclear, hadronic or plasma absorption at high $AB$, it can
be accommodated by reducing the value of
$\tau$.  What is shown here is that the quadratic gluon depletion effect by
itself is able to account for the enhanced suppression in the Pb-Pb data.

A concomitant phenomenon associated with quadratic gluon depletion is
the suppression of back-to-back $D\bar D$ production in Pb-Pb
collision, but not in
$AB$ collisions where $A$ is smaller.  Photon production would not
necessarily be suppressed, since the quarks produced by gluon conversion
can carry on the
$\gamma$-producing subprocess without inhibition.  Dilepton production may
or may not be enhanced, depending on whether the extra quarks and
antiquarks are produced inside or outside the interaction region.  It is
therefore important that all those signatures should be examined
experimentally in the collisions of very heavy ions.

Whether or not the nonlinear gluon depletion process can wholly or
partially account for the enhanced $J/\psi$ suppression phenomenon,
what we have discovered here is that there is a whole class of parton
interactions whose role in heavy-ion collisions has hitherto been
overlooked, but they are of crucial importance to any process whose rate
depends on the magnitude of the gluon flux available in large nuclei.

This work was supported, in part, by the U.S.-Slovakia Science and
Technology Program, National Science Foundation under Grant No.
INT-9319091 and by the U. S. Department of Energy under Grant No.
DE-FG03-96ER40972.

\vspace{1cm}
\begin{center}
\section*{Figure Captions}
\end{center}
\begin{description}

\item[Fig.\ 1]\quad The suppression factor $S^{AB}_{J/\psi}$,
abbreviated as $S$, is plotted against $AB$, when only the quadratic
depletion effect is taken into account. The shaded regions are for
$\tau$ having values between 0.5 (upper boundaries) and 1.0 (lower
boundaries). Three values of $\sigma_{in}$ are used, as indicated.

\item[Fig.\ 2]\quad The suppression factor $S^{AB}_{J/\psi}$,
abbreviated as $S$, is plotted against $AB$, when both the usual linear
(mainly absorption) effect and the quadratic depletion effect are taken
into account. The data are from [4]. Typical values for the parameters in
the theoretical calculations have been used.
\end{description}


\vskip 1cm

\begin{thebibliography}{99}

\bibitem{hpp} R. C. Hwa, J. Pi\v{s}\'{u}t, and N. Pi\v{s}\'{u}tov\'a, OITS-621
(nucl-th/9702051), to be published in Phys. Rev. C.

\bibitem{for} For a review see H. Satz, and J.-P. Blaizot and J.-Y.
Ollitrault in
{\it Quark-Gluon Plasma}, edited by R.C. Hwa (World Scientific, Singapore,
1990), and D. Kharzeev and H. Satz in {\it Quark-Gluon Plasma 2}, edited
by R.C. Hwa (World Scientific, Singapore, 1995).

\bibitem{cl}  C. Louren\c{c}o,  Nucl.\ Phys.\ {\bf A610}, 552c
(1996).  

\bibitem{mg}M.\ Gonin {\it et al}., (NA50), Nucl.\ Phys.\ {\bf A610}, 404c
(1996).  

\bibitem{kl}D.\ Kharzeev, C.\ Louren\c{c}o, M.\ Nardi, and H.\
Satz, Z. Phys. C {\bf 74}, 307 (1997).
\bibitem{bo}J.-P.\ Blaizot and J.-Y.\ Ollitrault, Phys.\ Rev.\
Lett.\ {\bf 77}, 1703 (1996).

\bibitem{gv}S.\ Gavin and R.\ Vogt, Nucl.\ Phys.\ {\bf A610}, 442c (1996);
Phys.\ Rev.\ Lett.\ {\bf 78}, 1006 (1997).

\bibitem{ckkg}A.\ Capella, A.\ Kaidalov, A.\ Kouidu Akil, and C.\
Gerschel, Phys.\ Lett.\ {\bf B393}, 431 (1997).
\bibitem{fpp}J.\ Ft\'{a}\u{c}nik, J.\ Pi\v{s}\'{u}t, and N.\
Pi\v{s}\'{u}tov\'{a}, hep-ph/9604304.
\bibitem{ck}W.\ Cassing and C.\ M.\ Ko, Phys. Lett. {\bf B396}, 39 (1997).

\bibitem{wo}C.\ Y.\ Wong, Phys.\ Rev.\ C {\bf 55}, 2621 (1997); Phys.\
Rev.\ Lett.\ {\bf 76}, 196 (1996).

\bibitem{gh}C.\ Gerschel and J.\ H\"{u}fner,   Phys. Lett.
B {\bf 207}, 253 (1988);   Z.\ Phys.\ C
 {\bf 56}, 171 (1992).

\bibitem{hw}R.\ C.\ Hwa and X.\ -N.\ Wang,  Phys.\
Rev. D {\bf 39}, 2561 (1989).

\end{thebibliography}
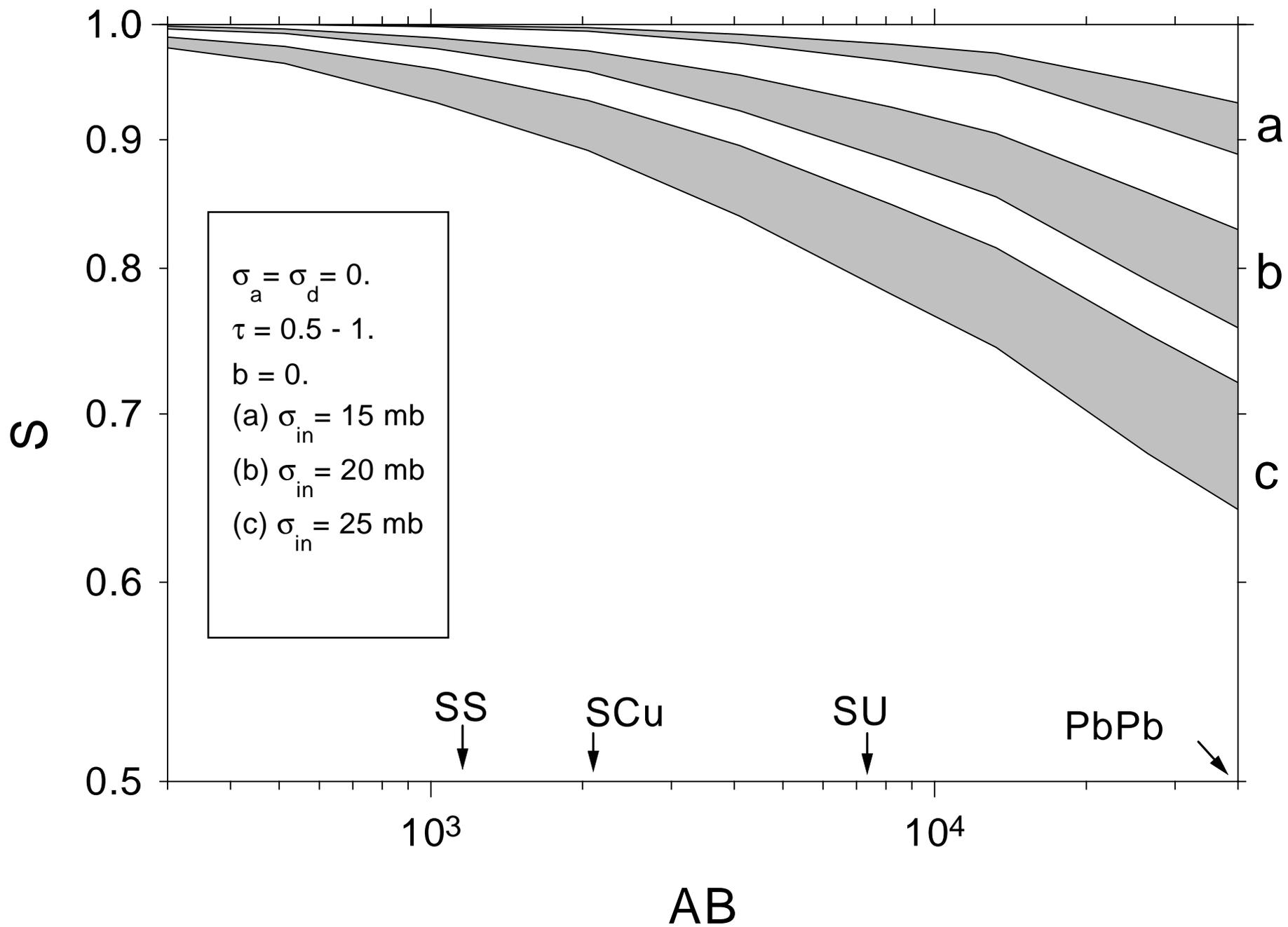
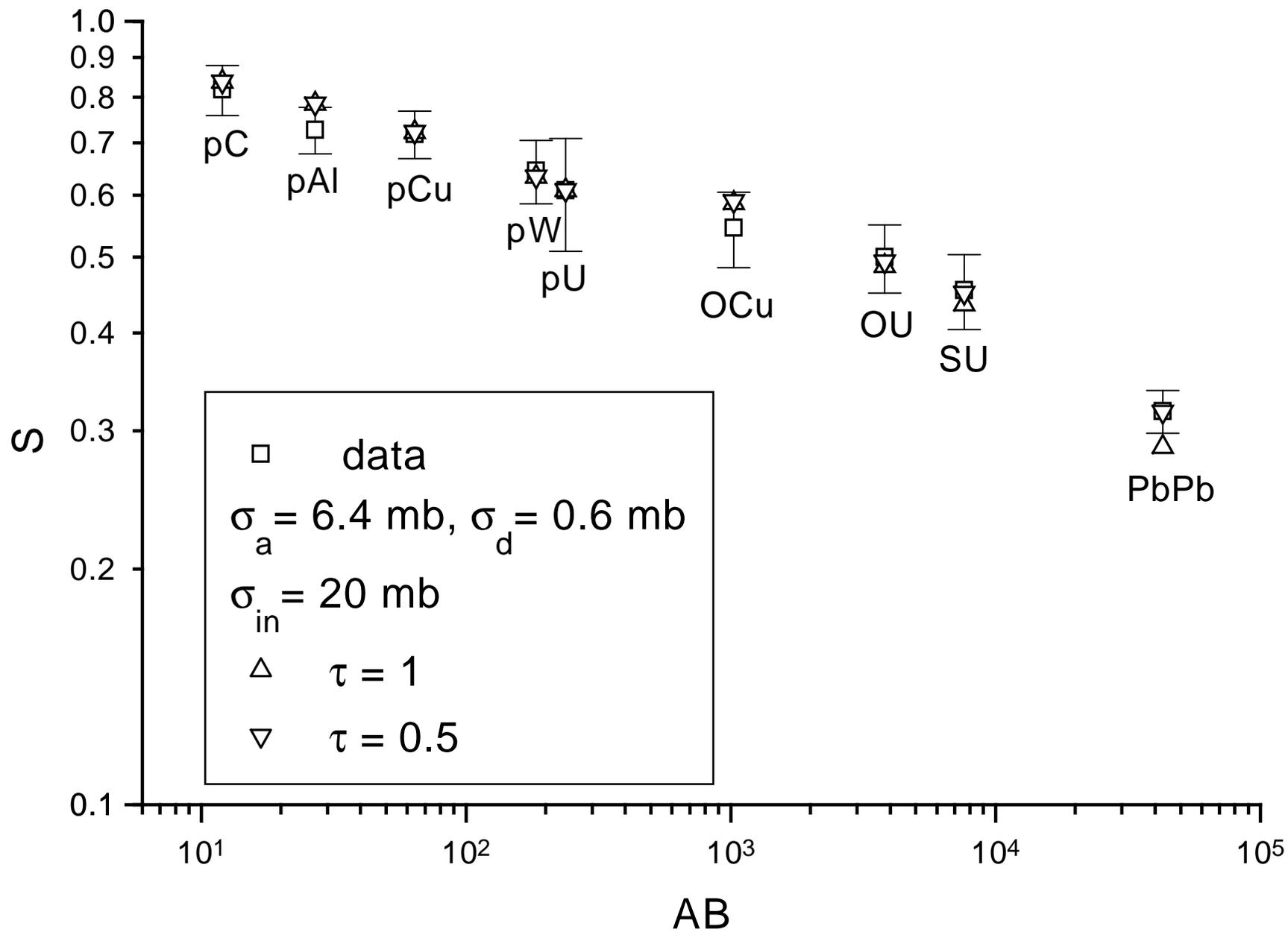
\end{document}